\documentclass{article}

\usepackage{arxiv}

\usepackage[utf8]{inputenc} 
\usepackage[T1]{fontenc}
\usepackage{hyperref}
\usepackage{url}
\usepackage{booktabs}
\usepackage{amsfonts}
\usepackage{amsmath}
\usepackage{nicefrac}
\usepackage{microtype}
\usepackage{lipsum}
\usepackage{graphicx}
\usepackage{natbib}
\usepackage{doi}

\title{TMMax: High-performance modeling of multilayer thin-film structures using transfer matrix method with JAX}

\author{ \href{https://orcid.org/0009-0002-9880-0446}{\includegraphics[scale=0.06]{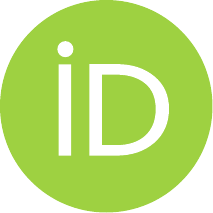}\hspace{1mm}Bahrem Serhat~Danis}\\
	Department of Electrical and Electronics Engineering\\
	Koç University\\
	Istanbul, 34450, Turkey \\
	\texttt{bdanis23@ku.edu.tr} \\
	\And
	\href{https://orcid.org/0000-0001-5887-0293}{\includegraphics[scale=0.06]{orcid.pdf}\hspace{1mm}Esra~Zayim} \\
	Physics Engineering Department\\
	Istanbul Technical University\\
	Istanbul, 34469, Turkey \\
}

\date{October 24, 2025}

\renewcommand{\headeright}{Technical Note - October 24, 2025}
\renewcommand{\undertitle}{Technical Note}
\renewcommand{\shorttitle}{}

\hypersetup{
pdftitle={TMMax: High-performance modeling of multilayer thin-film structures using transfer matrix method with JAX},
pdfsubject={physics.comp-ph, physics.optics},
pdfauthor={Bahrem Serhat ~Danis, Esra ~Zayim},
pdfkeywords={Multilayer Thin-Film, Transfer Matrix Method, Vectorization, Coating Design, JAX, Python, Optics},
}

\begin{document}
\maketitle

\begin{abstract}
Optical multilayer thin-films are fundamental components that enable the precise control of reflectance, transmittance, and phase shift in the design of photonic systems. Rapid and accessible simulation of these structures holds critical importance for designing and analyzing complex coatings. While researchers commonly use the traditional transfer matrix method for designing these structures, its scalar approach to wavelength and angle of incidence causes redundant recalculations and inefficiencies in large-scale simulations. Furthermore, traditional method implementations do not support automatic differentiation, which limits their applicability in gradient-based inverse design approaches. Here, we present TMMax, a Python library that fully vectorizes and accelerates transfer matrix method using the high-performance machine learning library JAX. TMMax supports CPU, GPU, and TPU hardware, and includes a publicly available material database. Our approach, demonstrated through benchmarking, allows us to model thin-film stacks with hundreds of layers within seconds. This illustrates that our method achieves a simulation speedup of x100s over a baseline NumPy implementation, enabling optical engineers and thin-film researchers in optics and photonics to efficiently design complex dielectric multilayer structures through rapid and scalable simulations.
\end{abstract}


\section{Statemement of need}
\label{sec:statemementofneed}

The Transfer Matrix Method (TMM) models multilayer optical thin films by applying Snell’s law for light propagation and Fresnel equations to compute interface transmittance and reflectance.

\vspace{1.5em}

\begin{equation}
\mathbf{M} = \prod_{i=0}^{N-2} \mathbf{M}_i
\label{eq:total_matrix}
\end{equation}

\vspace{1.5em}

In TMM, the optical behavior of an N-layer multilayer structure composed of dielectric materials is obtained by computing the system matrix $\mathbf{M}$, as shown in Equation~\eqref{eq:total_matrix}. This matrix calculation, commonly referred to as the Abeles TMM \cite{refId0}, results from the successive multiplication of the transfer matrices of each layer ($\mathbf{M}_i$) \cite{katsidis2002general}. 

\begin{figure}[!htbp]
\centering
\includegraphics[width=\textwidth,keepaspectratio]{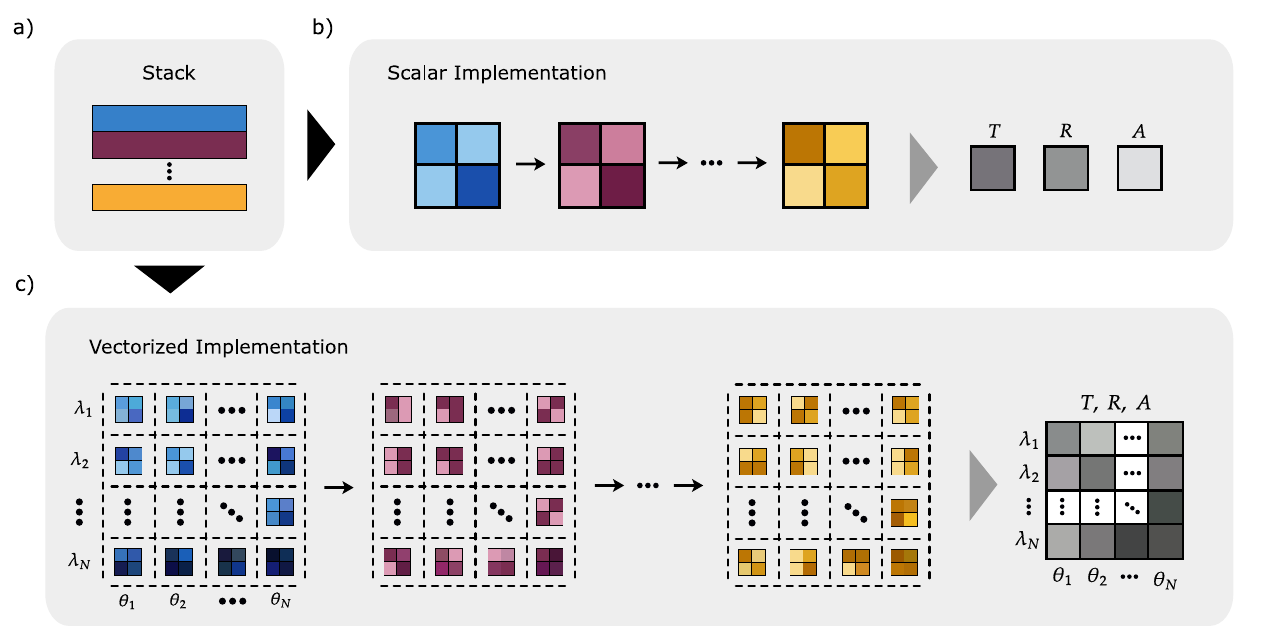}
\caption{Schematic of two strategies for calculating transmission, reflection, and absorption in multilayer thin-film simulations. The system (a) is modeled either by sequentially multiplying 2×2 transfer matrices for each wavelength and incidence angle (b) or by vectorizing these operations across both axes (c).}
\end{figure}

\newpage

In traditional TMM implementations, the stack of layers in Figure 1a is simulated using a single wavelength and angle of incidence, as shown in Figure 1b, and nested loops over wavelengths and angles lead to redundant calculations \cite{byrnes2020multilayeropticalcalculations}. TMMax removes these redundancies by vectorizing wavelengths and angles and all intermediate TMM operations via JAX \cite{jax2018github}. As seen in the schematic of the vectorized implementation in Figure 1c, we vectorize all intermediate operations in TMM and subsequently apply JAX’s just-in-time (JIT) decorator. Instead of running the mapped TMM code sequentially over each batch element of wavelength and angle of incidence, \texttt{jax.jit} fuses all operations across the batch into a single XLA-compiled \cite{xla2023github} kernel. This reduces function call overhead and provides a faster TMM implementation. TMMax replaces the conventional for-loop system-matrix calculation \cite{6131837} with JAX’s \texttt{lax.scan}, enabling JIT compilation and eliminating interpreter bottlenecks, while running efficiently on CPUs, GPUs, and TPUs without code changes. 

TMMax supports deep learning–based inverse design by keeping all computations on the GPU, avoiding costly CPU–GPU data transfers \cite{10.1117/1.OE.58.6.065103}. Whereas NumPy-based \cite{2020NumPy-Array} TMM packages that lack native gradients and require Autograd \cite{maclaurin2015autograd}, TMMax natively computes gradients. Additionally, TMMax integrates a curated database of 30 extensively used dielectric materials, sourced from refractiveindex.info \cite{polyanskiy2024refractiveindex}, thereby enabling optical engineers and thin-film researchers in optics and photonics to efficiently simulate complex multilayer structures through a scalable, JAX-accelerated implementation.

\section{Benchmarks}
\label{sec:benchmarks}

Runtime in TMM scales naturally with the number of layers, as well as the lengths of the wavelength and incidence-angle arrays, due to the increased number of transfer matrix multiplications. To benchmark TMMax, we used tmm library \cite{byrnes2020multilayeropticalcalculations} as a reference.

\begin{figure}[!htbp]
\centering
\includegraphics[width=0.75\textwidth]{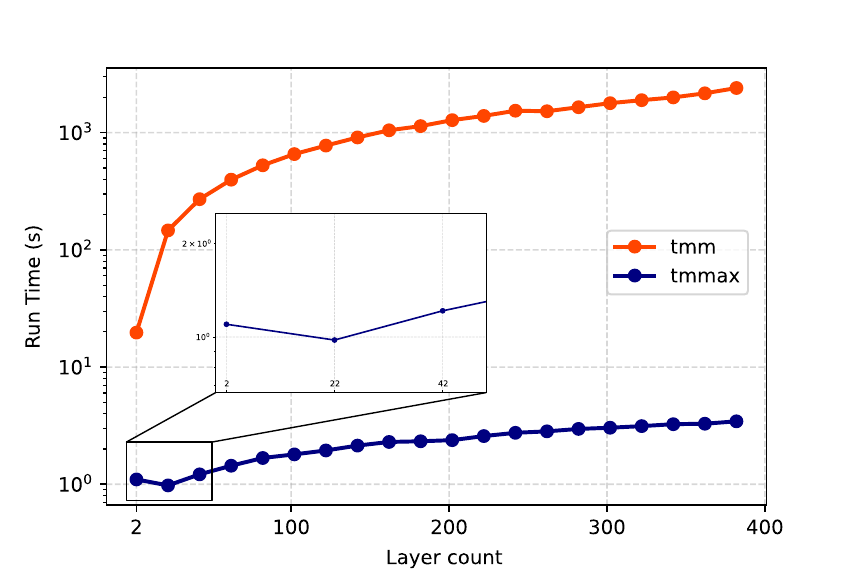}
\caption{Run time vs. layer count comparing tmm (orange) and TMMax (blue).}
\end{figure}

\newpage

To assess how layer count affects computational performance, we sampled 20 multilayer structures ranging from 2 to 400 layers, with each layer randomly assigned one of seven materials and thicknesses between 100–500 nm. Spectral and angular domains were fixed at 20 points each, spanning 500–1000 nm and 0-$\pi$/2 radians, respectively. Figure 2 shows that while tmm runtime grows rapidly, TMMax scales efficiently, remaining nearly constant ($\sim$1.0–1.2 s) for low-layer structures and achieving speedups from 18× (2 layers) to 700× (400 layers).

\begin{figure}[!htbp]
\centering
\includegraphics[width=0.75\textwidth]{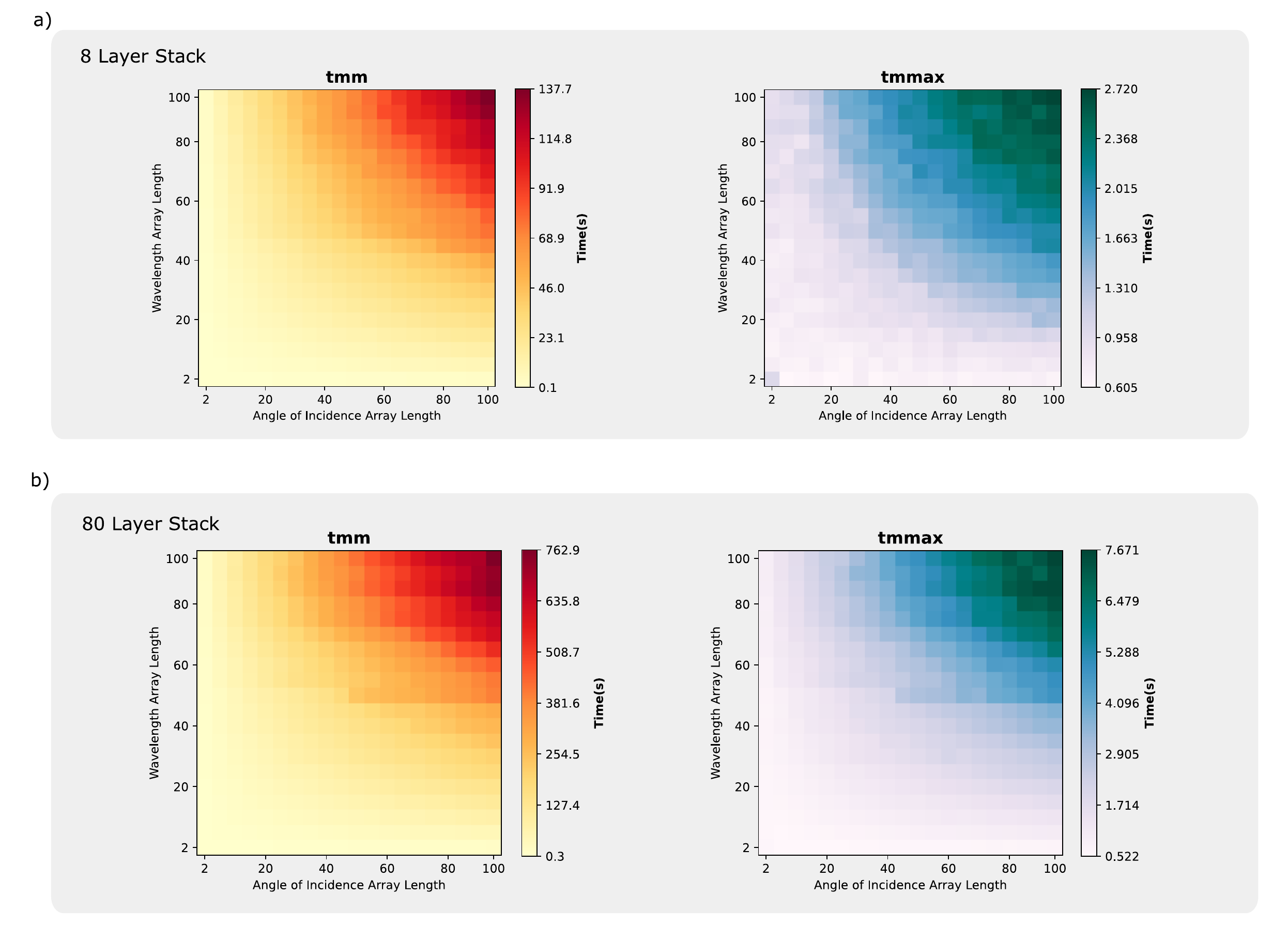}
\caption{The colormaps show the runtime performance of tmm and TMMax across varying simulation grid sizes, comparing 8- and 80-layer stacks in (a) and (b), respectively.}
\end{figure}

We benchmarked the effects of wavelength and incident angle array sizes by sampling 20 values from 2 to 100, generating simulation grids from 2×2 to 100×100 for an 8-layer structure (Figure 3a). tmm runtime rises sharply with grid size, reaching $\sim$138 s for 100×100, whereas TMMax remains below 3 s. For the smallest 2×2 grid, tmm is faster ($\sim$0.1 s vs. $\sim$0.6 s) due to NumPy’s low overhead, while JAX incurs higher initialization costs. As layers increase to 80 (Figure 3b), tmm exceeds 760 s, but TMMax stays under 8 s, demonstrating superior efficiency and stability against both problem size and structural complexity.

We used Python’s timeit module to benchmark each simulation 50 times, with all comparisons run on a single Intel Core i9 core without GPU or multicore use for fairness.

\section{Installation}
\label{sec:installation}

TMMax can be readily installed from the Python Package Index using `pip install tmmax`, which automatically handles all dependencies. For detailed installation instructions and platform compatibility, please refer to the \href{https://tmmax.readthedocs.io/en/latest/index.html}{TMMax Documentation}.

\section{Acknowledgements}
\label{sec:acknowledgements}

This work was supported by the Scientific and Technological Research Council of Türkiye (TUBITAK) under the 2209-A Research Project Support Programme for Undergraduate Students, 2022 First-Term Call.

\bibliographystyle{unsrt}
\bibliography{references}

\begin{thebibliography}{10}

\bibitem{refId0}
{Abelès, Florin}.
\newblock Recherches sur la propagation des ondes électromagnétiques
  sinusoïdales dans les milieux stratifiés - application aux couches minces.
\newblock {\em Ann. Phys.}, 12(5):596--640, 1950.

\bibitem{katsidis2002general}
Charalambos~C Katsidis and Dimitrios~I Siapkas.
\newblock General transfer-matrix method for optical multilayer systems with
  coherent, partially coherent, and incoherent interference.
\newblock {\em Applied Optics}, 41(19):3978--3987, 2002.

\bibitem{byrnes2020multilayeropticalcalculations}
Steven~J. Byrnes.
\newblock Multilayer optical calculations, 2020.

\bibitem{jax2018github}
James Bradbury, Roy Frostig, Peter Hawkins, Matthew~James Johnson, Chris Leary,
  Dougal Maclaurin, George Necula, Adam Paszke, Jake Vander{P}las, Skye
  Wanderman-{M}ilne, and Qiao Zhang.
\newblock {JAX}: composable transformations of {P}ython+{N}um{P}y programs,
  2018.

\bibitem{xla2023github}
{OpenXLA Team}.
\newblock {XLA: Accelerated Linear Algebra Compiler for Machine Learning}.
\newblock \url{https://github.com/openxla/xla}, 2023.
\newblock Accessed: 2025-07-13.

\bibitem{6131837}
Kazufumi Nishida, Yasuaki Ito, and Koji Nakano.
\newblock Accelerating the dynamic programming for the matrix chain product on
  the gpu.
\newblock In {\em 2011 Second International Conference on Networking and
  Computing}, pages 320--326, 2011.

\bibitem{10.1117/1.OE.58.6.065103}
Ravi~S. Hegde.
\newblock {Accelerating optics design optimizations with deep learning}.
\newblock {\em Optical Engineering}, 58(6):065103, 2019.

\bibitem{2020NumPy-Array}
Charles~R. Harris, K.~Jarrod Millman, Stéfan~J van~der Walt, Ralf Gommers,
  Pauli Virtanen, David Cournapeau, Eric Wieser, Julian Taylor, Sebastian Berg,
  Nathaniel~J. Smith, Robert Kern, Matti Picus, Stephan Hoyer, Marten~H. van
  Kerkwijk, Matthew Brett, Allan Haldane, Jaime Fernández~del Río, Mark
  Wiebe, Pearu Peterson, Pierre Gérard-Marchant, Kevin Sheppard, Tyler Reddy,
  Warren Weckesser, Hameer Abbasi, Christoph Gohlke, and Travis~E. Oliphant.
\newblock Array programming with {NumPy}.
\newblock {\em Nature}, 585:357–362, 2020.

\bibitem{maclaurin2015autograd}
Dougal Maclaurin, David Duvenaud, and Ryan~P Adams.
\newblock Autograd: Effortless gradients in numpy.
\newblock In {\em ICML 2015 AutoML workshop}, volume 238. CNRS, 2015.

\bibitem{polyanskiy2024refractiveindex}
Mikhail~N. Polyanskiy.
\newblock Refractiveindex.info database of optical constants.
\newblock {\em Scientific Data}, 11(1):94, 2024.

\end{thebibliography}

\end{document}